\documentclass[nohyperref]{article}

\usepackage{microtype}
\usepackage{graphicx}
\usepackage{subfigure}
\usepackage{booktabs} 

\usepackage{amsfonts}
\usepackage{url}

\usepackage{hyperref}

\usepackage{algorithm}

\usepackage[accepted]{icml2022}

\usepackage{amsmath}
\usepackage{amssymb}
\usepackage{mathtools}
\usepackage{amsthm}

\usepackage{multirow}

\usepackage[capitalize,noabbrev]{cleveref}

\makeatletter
\newcommand{\thickhline}{%
	\noalign {\ifnum 0=`}\fi \hrule height 1pt
	\futurelet \reserved@a \@xhline
}

\theoremstyle{plain}

\theoremstyle{definition}

\theoremstyle{remark}

\usepackage[textsize=tiny]{todonotes}

\icmltitlerunning{DiffGAN-TTS: High-Fidelity and Efficient Text-to-Speech with Denoising Diffusion GANs}

\begin{document}


\twocolumn[
\icmltitle{DiffGAN-TTS: High-Fidelity and Efficient Text-to-Speech \\ with Denoising Diffusion GANs}



\icmlsetsymbol{equal}{*}

\begin{icmlauthorlist}
\icmlauthor{Songxiang Liu}{1}
\icmlauthor{Dan Su}{1}
\icmlauthor{Dong Yu}{1}
\end{icmlauthorlist}

\icmlaffiliation{1}{Tencent AI Lab}

\icmlcorrespondingauthor{Songxiang Liu}{songxiangliu.cuhk@gmail.com}

\icmlkeywords{Machine Learning, ICML}

\vskip 0.3in
]



\printAffiliationsAndNotice{}  

\begin{abstract}

Denoising diffusion probabilistic models (DDPMs) are expressive generative models that have been used to solve a variety of speech synthesis problems. However, because of their high sampling costs, DDPMs are difficult to use in real-time speech processing applications.
In this paper, we introduce DiffGAN-TTS, a novel DDPM-based text-to-speech (TTS) model achieving high-fidelity and efficient speech synthesis.
DiffGAN-TTS is based on denoising diffusion generative adversarial networks (GANs), which adopt an adversarially-trained expressive model to approximate the denoising distribution.
We show with multi-speaker TTS experiments that DiffGAN-TTS can generate high-fidelity speech samples within only 4 denoising steps.
We present an active shallow diffusion mechanism to further speed up inference. A two-stage training scheme is proposed, with a basic TTS acoustic model trained at stage one providing valuable prior information for a DDPM trained at stage two.
Our experiments show that DiffGAN-TTS can achieve high synthesis performance with only 1 denoising step.

\end{abstract}

\section{Introduction}
\label{sec1:intro} 

Text-to-speech (TTS) synthesis is a typical multimodal generation task, where there could be various speech outputs (e.g., with different speaker identities, emotions, speaking styles, etc.) for a given text input. Typical modern neural TTS systems consist of three key components: text analysis frontend, acoustic model, and vocoder. The text analysis frontend normalizes input text and transforms it into linguistic representations. The acoustic model then converts the linguistic representation into time-frequency domain acoustic features, such as mel spectrograms. Finally, the vocoder module generates time-domain waveforms from acoustic features.
Different types of generative models have been used as acoustic models to model acoustic variation information in this one-to-many mapping problem to improve the expressiveness and fidelity of synthetic speech.

Neural network-based autoregressive (AR) models have been adopted for TTS and have shown the capability of generating highly natural speech by producing acoustic features frame by frame \cite{tacotron,deepvoice2,char2wav,transformer_tts}. Nonetheless, AR TTS models often suffer from pronunciation issues, e.g., word skipping and repeating, due to accumulated prediction errors at inference. Moreover, their sequential generative process limits the synthesis speed.
To address these limitations, various non-AR TTS models have been proposed. They either leverage an external text-to-acoustic alignment module \cite{fastspeech,fastspeech2,paranet,parallel_tacotron} or jointly train one within the TTS model \cite{aligntts,efficient-tts,badlani2021one}.
Other generative models have also been studied for TTS, such as Flow-based models\cite{glow-tts,flowtts}, variational auto-encoder (VAE)-based models\cite{bvae-tts,liu2021vara}, and generative adversarial network (GAN)-based models \cite{EATS,ganspeech}. TTS models combining different generative modeling techniques are also investigated, such as Flow with VAE\cite{portaspeech}, Flow with VAE and GAN \cite{vits}.

\begin{figure*}[t]
	\centering
	\includegraphics[width=12cm]{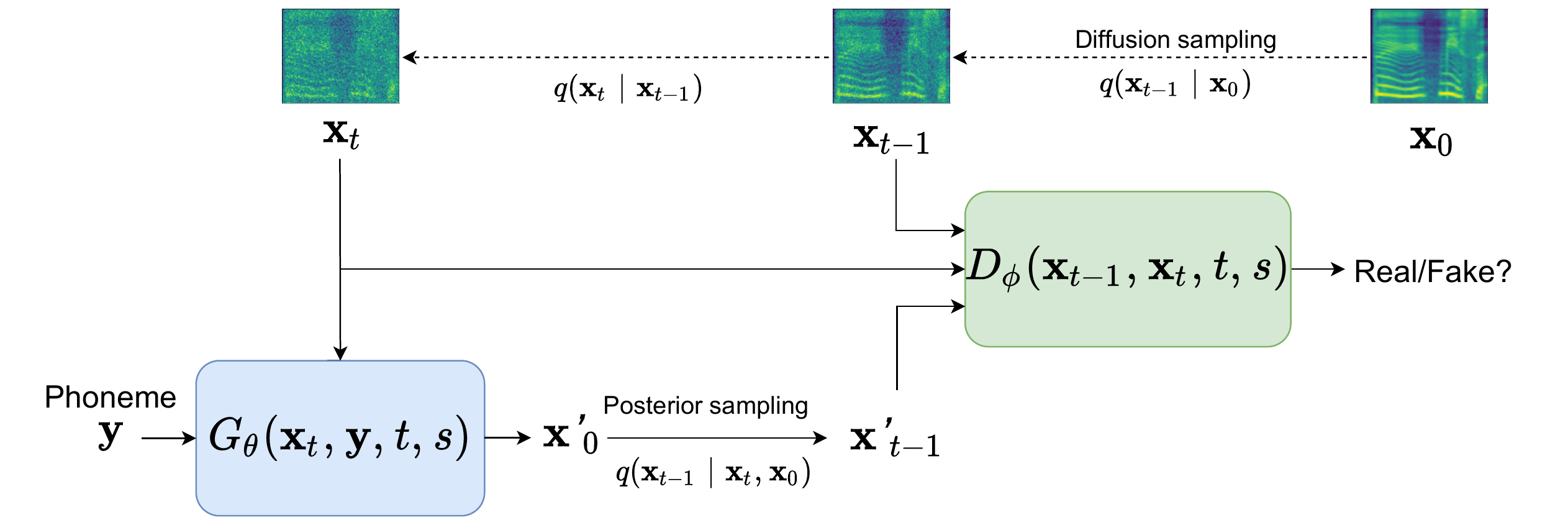}				
	\caption{Training process of DiffGAN-TTS.}
	\label{fig:training}
\end{figure*}

Another class of generative models called denoising diffusion probabilistic models (DDPMs), or abbreviated as diffusion models, has shown an impressive capability to model complex data distributions. Diffusion models have obtained state-of-the-art results in several important domains, including image synthesis\cite{ddpm}, audio synthesis\cite{diffwave,wavegrad,grad-tts}, graphs\cite{niu2020permutation}, and symbolic music generation \cite{mittal2021symbolic}.
Typical diffusion models comprise a parameter-free $T$-step Markov chain called \textit{the diffusion process}, which gradually adds small random noise into the data, and a parameterized $T$-step Markov chain called \textit{the denoising process} (also known as \textit{the reverse process}), which removes the added noise as a denoising function.
In spite of wide and successful applications of diffusion models in different tasks, sampling from them is not efficient and often requires hundreds or even thousands of denoising steps, making them unsuited for real-time applications.
In \cite{diffgan}, this slow sampling issue of diffusion models is attributed to the fact that they commonly assume that the denoising distribution can be approximated by Gaussian distributions. This assumption imposes a constraint on typical diffusion models that the denoising step size is sufficiently small and the number of diffusion steps large enough.
To use larger denoising step sizes, a conditional GAN is adopted as a non-Gaussian multimodal function to model the denoising distribution \cite{diffgan}, leading to much more efficient sampling and, at the same time, competitive sample diversity and fidelity.

This paper introduces DiffGAN-TTS, which is a novel non-AR TTS model based on diffusion models and achieves high-fidelity and efficient TTS.
DiffGAN-TTS leverages the powerful modeling capability of diffusion models to address the challenging one-to-many text-to-spectrogram mapping problem.
Partially inspired by the denoising diffusion GAN model \cite{diffgan}, we model the denoising distribution with an expressive acoustic generator, which is adversarially trained to match the true denoising distribution. DiffGAN-TTS allows large denoising steps at inference, which greatly reduces the number of denoising steps and accelerates sampling.
We introduce an active shallow diffusion mechanism into DiffGAN-TTS to further accelerate its sampling process. A two-stage training scheme is designed, where a basic acoustic model trained in stage 1 provides strong prior information for a denoising model trained in stage 2.


\begin{figure*}[t]
	\centering
	\includegraphics[width=12cm]{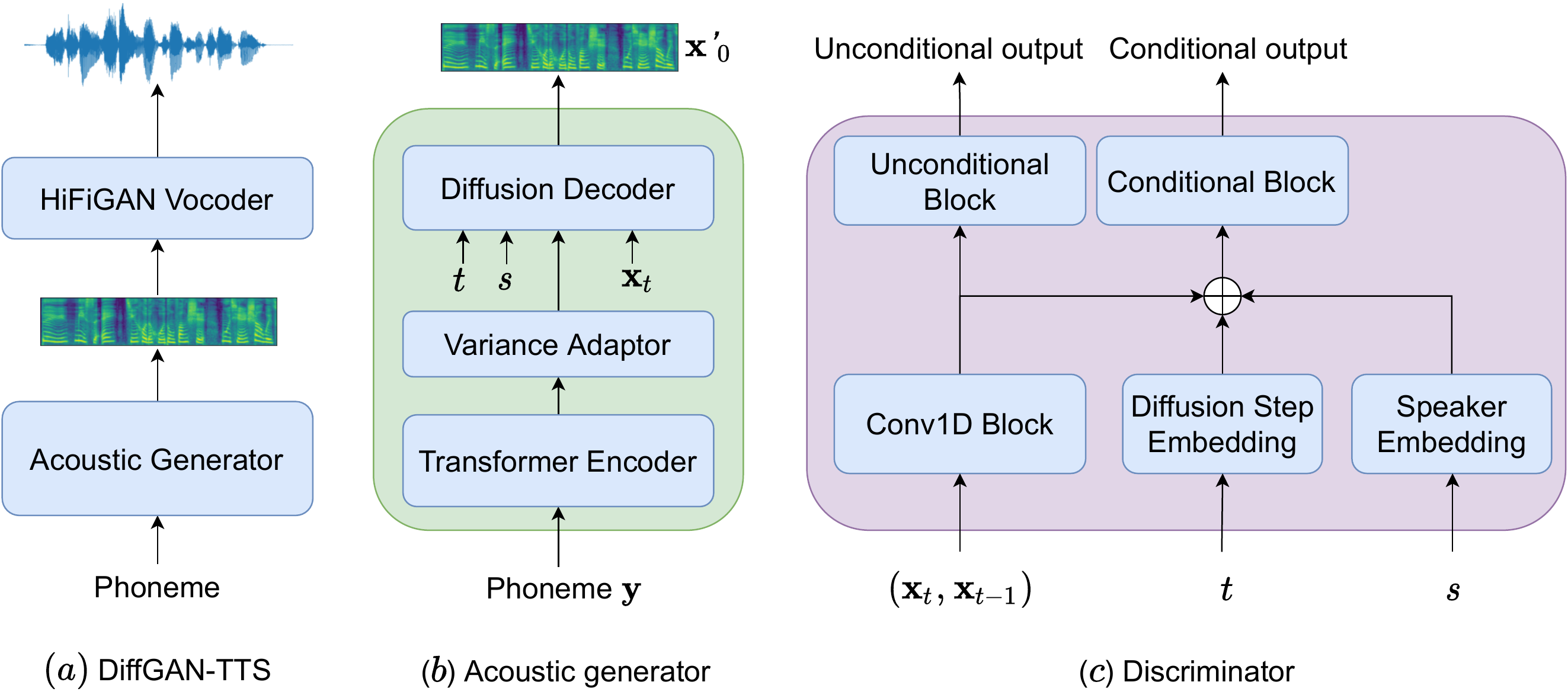}				
	\caption{The overall architecture of DiffGAN-TTS.}
	\label{fig:model_arc}
\end{figure*}

\section{Diffusion Models}
\label{sec:diff_model}

Diffusion models usually consist of a parameter-free ${T}$-step Markov chain named \textit{the diffusion process} and a parameterized $T$-step Markov chain called \textit{the reverse process} or \textit{the denoising process}. The diffusion process gradually adds small Gaussian noises into the data until the data structure is totally destroyed at step $T$, while the reverse process learns a denoising function to remove the added noise to restore the data structure.

We define $q(\mathbf{x}_0)$ as the data distribution on $\mathbb{R}^L$, where $L$ is the data dimension, and $q(\mathbf{x}_T)\sim\mathcal{N}(\mathbf{0},\mathbf{I})$ as the latent variable at step $T$. Let $\mathbf{x}_t\in\mathbb{R}^L$ for $t=0,1,\cdots,T$ be sequence of variables with the same dimension, where $t$ is the index for diffusion steps. The diffusion process is modeled as a Gaussian transformation chain from data $\mathbf{x}_0$ to the latent variable $\mathbf{x}_T$ with pre-defined variance schedule $\beta_1, \cdots,\beta_T$:
\begin{equation}
    q(\mathbf{x}_{1:T}|\mathbf{x}_0)=\prod_{t\geq 1}q(\mathbf{x}_t|\mathbf{x}_{t-1}),
\end{equation}
where $q(\mathbf{x}_t|\mathbf{x}_{t-1}):=\mathcal{N}(\mathbf{x}_t;\sqrt{1-\beta_t}\mathbf{x}_{t-1},\beta_t\mathbf{I}).$
The reverse or denoising process parameterzied with $\theta$ is defined by:
\begin{equation}
    p_\theta(\mathbf{x}_{0:T})=p(\mathbf{x}_T)\prod_{t\geq 1}p_\theta(\mathbf{x}_{t-1}|\mathbf{x}_t).
\end{equation}
The denoising distribution $p_\theta(\mathbf{x}_{t-1}|\mathbf{x}_t)$ is often modeled as a conditional Gaussian distribution as $p_\theta(\mathbf{x}_{t-1}|\mathbf{x}_t):=\mathcal{N}(\mathbf{x}_{t-1};\mathbf{\mu}_\theta(\mathbf{x}_t), \sigma_t^2\mathbf{I})$, where $\mu_\theta(\mathbf{x}_t, t)$ and $\sigma_t^2\mathbf{I}$ are the mean and variance for the denoising model.
Given the parameterized reverse process with the well-learned parameter $\theta$, the sampling process (i.e., the generative process) is to first sample a Gaussian noise $\mathbf{x}_T \sim \mathcal{N}(\mathbf{0}, \mathbf{I})$, and then iteratively sample $\mathbf{x}_{t-1}\sim p_\theta(\mathbf{x}_{t-1}|\mathbf{x}_t)$ for $t=T, T-1, \cdots, 1$ along the reverse process, according to the so-called Langevin dynamics. The $\mathbf{x}_0$ is the generated data.

The likelihood $p_\theta(\mathbf{x}_0)=\int p_\theta(\mathbf{x}_{0:T})\text{d}\mathbf{x}_{1:T}$ is intractable.
Hence, the goal of training is to maximize its evidence lower bound (ELBO$\leq\log p_\theta(\mathbf{x}_0)$), which can be optimized to match the true denoising distribution $q(\mathbf{x}_{t-1}|\mathbf{x}_t)$ with the parameterized denoising model $p_\theta(\mathbf{x}_{t-1}|\mathbf{x}_t)$ with:
\begin{equation} \label{eq4}
    \text{ELBO} = \sum_{t\geq 1}\mathbb{E}_{q_({\mathbf{x}_t})}[D_{KL}(q(\mathbf{x}_{t-1}|\mathbf{x}_t)||p_\theta(\mathbf{x}_{t-1}|\mathbf{x}_t)] + c
\end{equation}
where $D_{KL}$ denotes the Kullback-Leibler (KL) divergence and $c$ contains constant terms that does not dependent on $\theta$. The KL divergence terms in Eq.~\ref{eq4} are generally intractable due to the unknown true denoising distritbuion $q(\mathbf{x}_{t-1}|\mathbf{x}_t)$.
Therefore, in \cite{ddpm} the step size $\beta_{t}$ in the variance schedule is assumed to be small and the number of denoising steps $T$ to be large enough such that both the diffusion and denoising processes have the same functional form (i.e., conditional Gaussian). Based on this, \cite{ddpm} shows a certain parameterization, transforming the ELBO optimization problem into a simple regression problem.

\section{DiffGAN-TTS}
\label{sec:diffgan_tts}

Although DDPMs have demonstrated their capability in modeling complex data distributions, their slow inference speed prevents them from being used in real-time applications.
This is due to the two key commonly-made assumptions in DDPMs \cite{diffgan}, as alluded to in Section~\ref{sec1:intro}: First, the denoising distribution $p_\theta(\mathbf{x}_{t-1}|\mathbf{x}_t)$ is modeled with a Gaussian distribution. Second, the number of denoising steps $T$ is often assumed to be large enough such that $\beta_t$ is small.
When the denoising step gets larger and the data distribution is non-Gaussian, the true denoising distribution becomes more complex and multimodal, and in such a case, adopting a parameterized Gaussian transformation to approximate the denoising distribution is insufficient for high-quality generation. Conditional GANs have been adopted to model the multimodal denoising distribution in image generation tasks \cite{diffgan}. In this section, we show in great detail how we apply this idea to efficient and high-fidelity multi-speaker TTS.

\subsection{Acoustic generator and Discriminator}

In this work, we focus on the more challenging multi-speaker TTS tasks, which is much harder than single-speaker TTS since acoustic variations are enlarged by data from different speakers.
As illustrated in Figure~\ref{fig:model_arc}$(a)$, DiffGAN-TTS takes phoneme sequence (denoted as $\mathbf{y}$) obtained from a text analysis tool as input to generate intermediate mel-spectrogram features $\mathbf{x}_0$ with a multi-speaker acoustic generator and then uses a HiFi-GAN-based neural vocoder \cite{hifigan} to produce time-domain waveforms. DDPMs are introduced into the acoustic generator to solve the ill-posed multimodal phoneme-to-mel-spectrogram mapping problem.

The training process of DiffGAN-TTS is illustrated in Figure~\ref{fig:training}. Our goal is to reduce the number of denoising steps $T$ (e.g., $T\leq4$) of DiffGAN-TTS such that its inference process is efficient, adequate for real-time speech processing applications without degrading the quality of the generated speech. We focus on discrete-time diffusion models, where denoising steps are large, and use a conditional GAN to model the denoising distribution. DiffTTS-GAN trains a conditional GAN-based acoustic generator $p_\theta(\mathbf{x}_{t-1}|\mathbf{x}_t)$ to approximate the true denoising distribution $q(\mathbf{x}_{t-1}|\mathbf{x}_t)$ with an adversarial loss that minimizes a divergence $D_{\text{adv}}$ per denoising step:
\begin{equation}
    \min_{\theta}\sum_{t\geq 1}\mathbb{E}_{q(\mathbf{x}_t)}[D_{\text{adv}}(q(\mathbf{x}_{t-1}|\mathbf{x}_t)||p_\theta(\mathbf{x}_{t-1}|\mathbf{x_t}))],
\end{equation}
where we adopt the least-squares GAN (LS-GAN) training formulation \cite{lsgan} to minimize $D_{\text{adv}}$ because of its various successful practices in audio generation domain \cite{melgan, hifigan, ganspeech, vits}.

Let us denote the speaker-ID as $s$. The discriminator is designed to be diffusion-step-dependent and speaker-aware to aid the generator to achieve high-fidelity multi-speaker speech generation, as illustrated in Figure~\ref{fig:model_arc}($c$).
The discriminator, denoted as $D_\phi(\mathbf{x}_{t-1}, \mathbf{x}_t, t, s)$ with learnable parameters $\phi$, is modeled as joint conditional and unconditional (JCU)\cite{Vocgan, ganspeech}. It not only outputs unconditional logits, but also conditional logits, where diffusion step embedding and speaker embedding are regarded as conditions.

We follow the same scheme in \cite{diffgan} to parameterize the denoising function as an implicit denoising model. Specifically, instead of directly modeling $p_\theta(\mathbf{x}_{t-1}|\mathbf{x}_t)$ by predicting $\mathbf{x}_{t-1}$ from $\mathbf{x}_t$, the denoising function is modeled as $p_\theta(\mathbf{x}_{t-1}|\mathbf{x}_t):=q(\mathbf{x}_{t-1}|\mathbf{x}_t, \mathbf{x}_0=f_\theta(\mathbf{x}_t, t))$, where $\mathbf{x}_0$ is predicted from diffused sample $\mathbf{x}_t$ using a diffusion function $f_\theta(\mathbf{x}_t, t)$ parameterized with $\theta$. During training, $\mathbf{x}'_{t-1}$ is sampled using the posterior distribution $q(\mathbf{x}'_{t-1}|\mathbf{x}'_0, \mathbf{x}_t)$, where $\mathbf{x}'_0$ is a predicted version of $\mathbf{x}_0$. The predicted tuple $(\mathbf{x}'_{t-1}, \mathbf{x}_t)$ is then fed into the JCU discriminator to compute the divergence $D_{\text{adv}}$ to the corresponding bonafide counterpart $(\mathbf{x}_{t-1}, \mathbf{x}_t)$. Different from \cite{diffgan}, we do not use the latent variable $z\sim\mathcal{N}(\mathbf{0}, \mathbf{I})$ as an input to the acoustic generator, since the diffusion decoder takes variance-adapted text encodings (happened in variance adaptor of the acoustic generator) and speaker-ID as auxiliary input.
To summarize, the implicit distribution $f_\theta(\mathbf{x}_t, t)$ is modeled with the acoustic generator, denoted as $G_\theta(\mathbf{x}_t, \mathbf{y}, t, s)$, which predicts $\mathbf{x}_0$ from $\mathbf{x}_t$ conditioned on phoneme input $\mathbf{y}$, diffusion step index $t$ and speaker ID $s$.

\subsection{Training loss}
\label{sec:loss}

The discriminator is trained to minimize the loss
\begin{equation} \label{eq:d-loss}
\begin{split}
    \mathcal{L}_D = \sum_{t\geq 1}\mathbb{E}_{q(\mathbf{x}_t)q(\mathbf{x}_{t-1}|\mathbf{x}_t)}[(D_\phi(\mathbf{x}_{t-1}, \mathbf{x}_t, t, s)-1)^2] \\ + \mathbb{E}_{p_\theta(\mathbf{x}_{t-1}|\mathbf{x}_t)}[D_\phi(\mathbf{x}_{t-1}, \mathbf{x}_t, t, s)^2].
\end{split}
\end{equation}

To train the acoustic generator, we also use the feature matching loss $\mathcal{L}_{fm}$, which learns similarity metric to discriminate real and fake data in the feature space \cite{fmloss}.
$\mathcal{L}_{fm}$ is computed by summing $l_1$ distances between every discriminator feature maps of real and generated samples:
\begin{equation}
\resizebox{0.48\textwidth}{!}{
    $\mathcal{L}_{fm} = \mathbb{E}_{q(\mathbf{x}_t)}[\sum_{i=1}^N||D_\phi^i(\mathbf{x}_{t-1}, \mathbf{x}_t, t, s)-D_\phi^i(\mathbf{x}'_{t-1}, \mathbf{x}_t, t, s)||_1]$,
}
\end{equation}
where $N$ is the total number of hidden layers in the discriminator.
Acoustic reconstruction loss is also used as additional loss to train the acoustic generator following FastSpeech2 \cite{fastspeech2}, as:
\begin{equation} \label{eq:reconst-loss}
\begin{split}
    \mathcal{L}_{recon} = \mathcal{L}_{mel}(\mathbf{x}_0, \mathbf{x}'_0) + \lambda_d\mathcal{L}_{duration}(\mathbf{d}, \hat{\mathbf{d}}) + \\ \lambda_p\mathcal{L}_{pitch}(\mathbf{p}, \hat{\mathbf{p}}) + \lambda_e\mathcal{L}_{energy}(\mathbf{e}, \hat{\mathbf{e}}),
\end{split}
\end{equation}
where $\mathbf{d}$, $\mathbf{p}$ and $\mathbf{e}$ are target duration, pitch and energy, respectively, and $\mathbf{\hat{d}}$, $\mathbf{\hat{p}}$ and $\mathbf{\hat{e}}$ are their corresponding predicted values. $\lambda_d$, $\lambda_p$ and $\lambda_e$ are loss weights, which are all set to be 0.1. $\mathcal{L}_{mel}$ uses MAE loss, while $\mathcal{L}_{duration}$, $\mathcal{L}_{pitch}$ and $\mathcal{L}_{energy}$ use MSE loss. In total, the acoustic generator is trained by minimizing:
\begin{equation} \label{eq:g-loss}
    \mathcal{L}_G = \mathcal{L}_{adv} + \mathcal{L}_{recon} + \lambda_{fm}\mathcal{L}_{fm},
\end{equation}
where
\begin{equation}
    \mathcal{L}_{adv}=\sum_{t\geq 1}\mathbb{E}_{q(\mathbf{x}_t)}\mathbb{E}_{p_\theta(\mathbf{x}_{t-1}|\mathbf{x}_t)}[(D_\phi(\mathbf{x}_{t-1}, \mathbf{x}_t, t, s)-1)^2],
\end{equation}
and $\lambda_{fm}$ is a dynamically scaled scalar computed as $\lambda_{fm}=\mathcal{L}_{recon}/\mathcal{L}_{fm}$ following \cite{ganspeech}.
Detailed training procedure as well as inference procedure is presented in Appendix~\ref{app:2}.

\subsection{Active shallow diffusion mechanism}
\label{sec:active}

\begin{figure}[t]
	\centering
	\includegraphics[width=6cm]{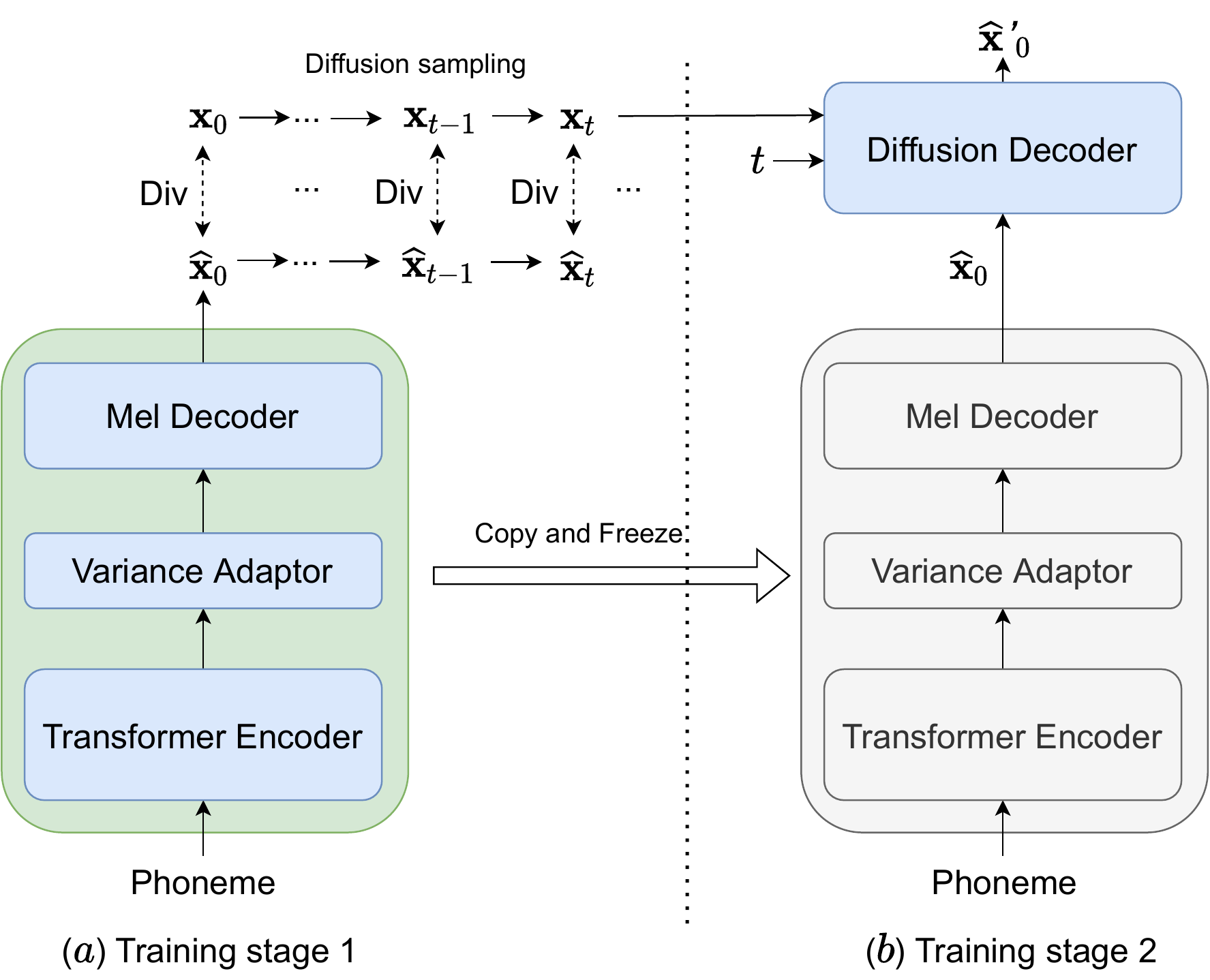}
	\caption{Two-stage training scheme.}
	\label{fig:two-stage}
\end{figure}

In the TTS literature, many acoustic models are trained with a simple loss, such as mean squared error (MSE) loss or mean absolute error (MAE) loss. Acoustic features generated from these acoustic models often suffer from over-smoothing issues due to incorrect uni-modal distribution assumptions on data, leading to non-desired synthesis performance. Nonetheless, the blurry acoustic predictions are not without use. It has been shown that output from acoustic models trained with MSE or MAE loss could provide strong prior knowledge of acoustic features (e.g., coarse harmonic structure), which could be further leveraged by a DDPM to generate refined features, leading to improved synthesis performance\cite{diffsinger}.

To further accelerate inference of DiffGAN-TTS, we introduce an active shallow diffusion mechanism.
As illustrated in Figure~\ref{fig:two-stage}, a two-stage training scheme is designed.
At training stage 1, a basic acoustic model, denoted as $G_{\psi}^{\text{base}}(\mathbf{y}, s)$ parameterized with $\psi$, is trained by:
\begin{equation} \label{eq:10}
    \min_\psi \sum_{t\geq 0}\mathbb{E}_{q(\mathbf{x}_t)}[\text{Div}(q_{\text{diff}}^t(G^{\text{base}}_{\psi}(\mathbf{y}, s)), q_{\text{diff}}^{t}(\mathbf{x}_0))],
\end{equation}
where $\text{Div}(\cdot, \cdot)$ is a distance function to measure the divergence between the predicted and ground-truth, and $q_{\text{diff}}^{t}(\cdot)$ is the diffusion sampling function at step $t$, e.g., $\mathbf{x}_t=q_{\text{diff}}^{t}(\mathbf{x}_0)$. It's worthy to note that $q_{\text{diff}}^{0}(\cdot)$ is an identity function. The training objective forces the base acoustic model to actively learn to make the diffused samples from ground-truth acoustic features and those from the predicted indistinguishable.
At training stage 2, pre-trained weights of the basic acoustic model are copied to initialize the corresponding weights of the acoustic generator of DiffGAN-TTS and then freeze, as illustrated in Figure~\ref{fig:two-stage}.
The base acoustic model generates coarse mel spectrogram $\hat{\mathbf{x}}_0$, which is taken as conditioning by the diffusion decoder.
The divergence $D_{\text{adv}}(q(\mathbf{x}_{t-1}|\mathbf{x}_t)||p_\theta(\mathbf{x}_{t-1}|\mathbf{x_t}))$ in Eq.~\ref{eq:10} is approximated by $D_{\text{adv}}(q(\mathbf{x}_{t-1}|\mathbf{x}_t)||p_\theta(\mathbf{x}_{t-1}|\mathbf{x}_{t}, \hat{\mathbf{x}}_0))$. Empirically, the diffusion decoder can be regarded as a post filter that conducts ``super-resolution" on the coarse prediction produced by the basic acoustic model.
We focus on reducing the number of denosing steps to \textbf{\textit{one}}. During inference, the basic acoustic model first generates a coarse mel spectrogram $\hat{\mathbf{x}}_0$, on which a diffused sample at diffusion step 1 is computed (i.e., $\hat{\mathbf{x}}_1$). Then DiffGAN-TTS takes $\hat{\mathbf{x}}_1$ as prior and runs one denoising step to get the final output. We term this variant as \textit{DiffGAN-TTS (two-stage)}, whose detailed training and inference procedures are presented in Appendix~\ref{app:2}. 

\begin{table*}[]
\caption{Objective and subjective evaluation as well as model efficiency results. SSIM, MCD, $F_0$ RMSE, and voice Cosine similarity (Cos. Sim.) are adopted as objective metrics. MOS, as the subjective metric, is presented with 95\% confidence intervals. RTFs are estimated on an NVIDIA T4 GPU.}
\vspace{0.3cm}
\centering
\begin{tabular}{lccccccc}
\thickhline
Model                   & SSIM($\uparrow$)  & MCD$_{24}$($\downarrow$)  & F$_0$ RMSE($\downarrow$) & Cos. Sim.($\uparrow$) & \#Params. & RTF    & MOS($\uparrow$) \\ \hline
GT                      & \multicolumn{6}{c}{-}                                  &  4.73$\pm$0.07   \\
GT (Mel+HiFi-GAN)       & 0.823 & 3.37 & 37.30    & 0.907     & 12.91M   & 0.0313 &  4.36$\pm$0.07   \\ \hline
FastSpeech 2            & 0.501 & 5.97 & 46.61   & 0.726     & 25.54M   & 0.0058 &  4.03$\pm$0.08   \\
GANSpeech               & 0.516 & 5.27 & 45.83   & 0.789     & 25.54M   & 0.0058 &  4.12$\pm$0.11   \\
DiffSpeech              & 0.498 & 5.58 & 48.77   & 0.736     & 44.43M   & 0.2224 &  \textbf{4.28$\pm$0.08}   \\ \hline
DiffGAN-TTS ($T$=1)       & 0.529 & 5.08 & 48.45   & 0.819     & 32.81M   & 0.0069 &  3.97$\pm$0.10   \\
DiffGAN-TTS ($T$=2)       & 0.524 & 5.06 & 47.24   & \textbf{0.823}     & 32.81M   & 0.0105 & 4.01$\pm$0.09    \\
DiffGAN-TTS ($T$=4)       & \textbf{0.532} & \textbf{4.93} & \textbf{45.68}   & 0.806     & 32.81M   & 0.0176 & \textbf{4.22$\pm$0.08}    \\
DiffGAN-TTS (two-stage) & 0.531 & 5.09 & 46.26   & 0.783     & 42.64M   & 0.0097 &  4.17$\pm$0.08   \\ \thickhline
\end{tabular} \label{tab:1}
\end{table*}

\subsection{Model architecture}

The transformer encoder in the acoustic generator of DiffGAN-TTS uses the same architecture as that in FastSpeech 2, which consists of 4 feed-forward transformer (FFT) blocks. The hidden size, number of attention heads, kernel size and filter size of the one-dimensional convolution in the FFT block are set as 256, 2, 9 and 1024, respectively. The variance adaptor has the same network structure and hyper-parameters as that in FastSpeech2, which consists of a duration predictor, a pitch predictor and an energy predictor. Differently, the pitch predictor and the energy predictor output phoneme-level fundamental frequency ($F_0$) contour and energy contour, respectively, whose labels are obtained by averaging frame-level $F_0$ and energy values according to phoneme-audio alignment information obtained from a hidden-Markov-model (HMM)-based forced aligner.

The diffusion decoder in DiffGAN-TTS uses a non-causal WaveNet architecture \cite{wavenet} with a slight modification.
We use a dilation rate of 1 because we are working with mel spectrograms rather than raw waveforms. With a dilation rate of 1, the diffusion decoder's receptive field is large enough. 
The diffusion decoder first performs an one-dimensional (1D) convolutional operation with kernel-size 1 (Conv1x1) on the noisy mel spectrogram $\mathbf{x}_t$ before applying the ReLU activation to the output.
Diffusion step $t$ is encoded using the same sinusoidal positional encoding as in \cite{vaswani2017attention}.
The mel spectrogram feature maps are added with the diffusion step embedding, which is then fed into 20 WaveNet residual blocks with a hidden dimension of 256.
The transformer encoder's output is imported into each residual block via separate Conv1x1 layers, whose output is added to the hidden feature maps. Then the gated mechanism introduced in \cite{wavenet} is used to further process the feature maps.
We add the skip connections from all WaveNet blocks and then process them with two Conv1x1 layers interleaved with a ReLU activation to get the diffusion decoder output. All WaveNet residual blocks have their speaker-IDs transformed into embedding vectors. Appendix~\ref{app:1} contains additional information.

The JCU discriminator, as depicted in Figure~\ref{fig:model_arc}($c$), adopts purely convolution networks. The Conv1D block consists of 3 one-dimensional convolutional layers with LeakyReLU (slope=$0.2$) as the activation function. The diffusion step embedding layer is the same as that in the diffusion decoder introduced above. The unconditional block and the conditional block have the same network structure, consisting of two 1D convolutional layers. The channels of convolution are 64, 128, 512, 128, and 1. The kernel sizes are 3, 5, 5, 5, and 3 and the strides are 1, 2, 2, 1, and 1.

The transformer encoder and variance adaptor in the basic acoustic model introduced in Section~\ref{sec:active} are the same as those in DiffGAN-TTS acoustic generator. The mel decoder is composed of 4 FFT blocks.

\section{Experiments}
\label{sec:exp}

\subsection{Data and Preprocessing}

We conduct experiments on an internal gender-balanced corpus comprising transcribed speech data from 228 Mandarin Chinese speakers.
In total, the corpus has 200 hours of speech data.
We randomly split 1024 utterances for validation and another 1024 utterances for testing. Speech samples are all sampled at 24,000 Hz with 16-bit quantization. Mel spectrograms with 80 frequency bins are computed through a short-time Fourier transform (STFT) using a 1024-point window size and a 10 ms frame-shift. We use the PyWorld toolkit \footnote{\url{https://github.com/JeremyCCHsu/Python-Wrapper-for-World-Vocoder}} to compute $F_0$ values from speech signals. Energy features are computed by taking the $l_2$-norm of frequency bins in STFT magnitudes.


\subsection{Training}

We train DiffGAN-TTS models with $T$=1, 2, and 4 using the Adam optimizer \cite{kingma2014adam}, with $\beta_1=0.5$ and $\beta_2=0.9$, for both the generator and the discriminator. The detailed computation of the variance schedule is presented in Appendix~\ref{app:2}. We use an exponential learning rate decay with rate 0.999 for training both the generator and discriminator. Initial learning rate for the generator is $10^{-4}$, while that for the discriminator is $2\times 10^{-4}$. Models are trained using one NVIDIA V100 GPU. The batch size is set to 64, and models are trained for at least 300k steps until losses converge. In the two-stage training scheme, the basic acoustic model is trained 200k steps with the Adam optimizer with $\beta_1=0.9$ and $\beta_2=0.98$ and follows the same learning rate schedule in \cite{vaswani2017attention}. As the distance function, we employ the simple MAE loss.

\subsection{Experimental Setup for Comparison}

We compare the proposed DiffGAN-TTS model with three strong counterpart models. The first counterpart is the representative non-AR TTS model FastSpeech 2 \cite{fastspeech2}. The second model is the GANSpeech model introduced in \cite{ganspeech}. The third model is the DiffSpeech model presented in \cite{diffsinger}. We use 60 denoising steps for the best performance on our data.

All compared models use the same acoustic feature setting and the same HiFi-GAN vocoder to ensure fairness. The official implementation\footnote{\url{https://github.com/jik876/hifi-gan}} (the ``config\_v1.json'' configuration) is used with slight changes. Since we use 24 kHz audio samples and the hop-size for computing the mel spectrogram is 240, we factorize the upsample rate as $240=8\times5\times3\times2$.
Moreover, we use the temporal nearest interpolation layer followed by a 1D convolutional layer as the upsampling operation to avoid possible checkerboard artifacts caused by the "ConvTranspose1d" upsampling layer \cite{odena2016deconvolution}.

\section{Results}
\label{sec:results}

\subsection{Objective Evaluation}

The quality and voice similarity of generated speech are measured.
We conducted an objective evaluation using the structural similarity index measure (SSIM) \cite{ssim}, mel-cepstral distortion (MCD)\cite{mcd}, $F_0$ root mean squared error (RMSE), and voice Cosine similarity. The computation of MCD and $F_0$ adopts dynamic time warping (DTW) \cite{dtw} to align the generated speech and the corresponding ground-truth recording. We use the first 24 coefficients when computing mel cepstrums. For computing voice Cosine similarity, we use a pre-trained speaker classifier to extract embedding vectors (the so-called \textit{d-vectors}) of a generated sample and its ground-truth counterpart, and then compute the Cosine distance between the two vectors.
The results are shown in Table~\ref{tab:1}.
We have the following observations:
1) The four DiffGAN-TTS models achieve the best SSIM, MCD, $F_0$ RMSE and voice Cosine similarity. Specifically, the DiffGAN-TTS ($T$=4) model outperforms other compared models in terms of SSIM, MCD, and $F_0$ RMSE. This indicates that adopting adversarial training in diffusion models well addresses the one-to-many mapping problem in TTS and is able to achieve high-quality speech synthesis, even when modeling multiple speakers within one model.
2) The best voice Cosine similarity is achieved by the proposed DiffGAN-TTS ($T$=2) model, and meanwhile, DiffGAN-TTS ($T$=1) has the second-highest voice Cosine similarity score. Although the voice Cosine similarity of DiffGAN-TTS ($T$=4) is not the best, it reaches a value of up to 0.806, which is still better than previous TTS models (e.g., FastSpeech 2, GANSpeech and DiffSpeech). Comparing the voice Cosine similarity of DiffGAN-TTS ($T$=1, 2, 4 and two-stage) with that of DiffSpeech, which takes up to 60 diffusion steps at inference, we conjecture that taking more diffusion steps should degrade voice similarity.
3) The DiffGAN-TTS (two-stage) model, which uses the active shallow diffusion mechanism, achieves overall good performance in terms of the four objective metrics. It is noteworthy that DiffGAN-TTS (two-stage) obtains better SSIM and $F_0$ RMSE results than DiffGAN-TTS ($T$=1 and $T$=2) and an on-par MCD score, indicating the effectiveness of the proposed active shallow diffusion mechanism.

\subsection{Subjective Evaluation}

\begin{figure}[t]
	\centering
	\includegraphics[width=6cm]{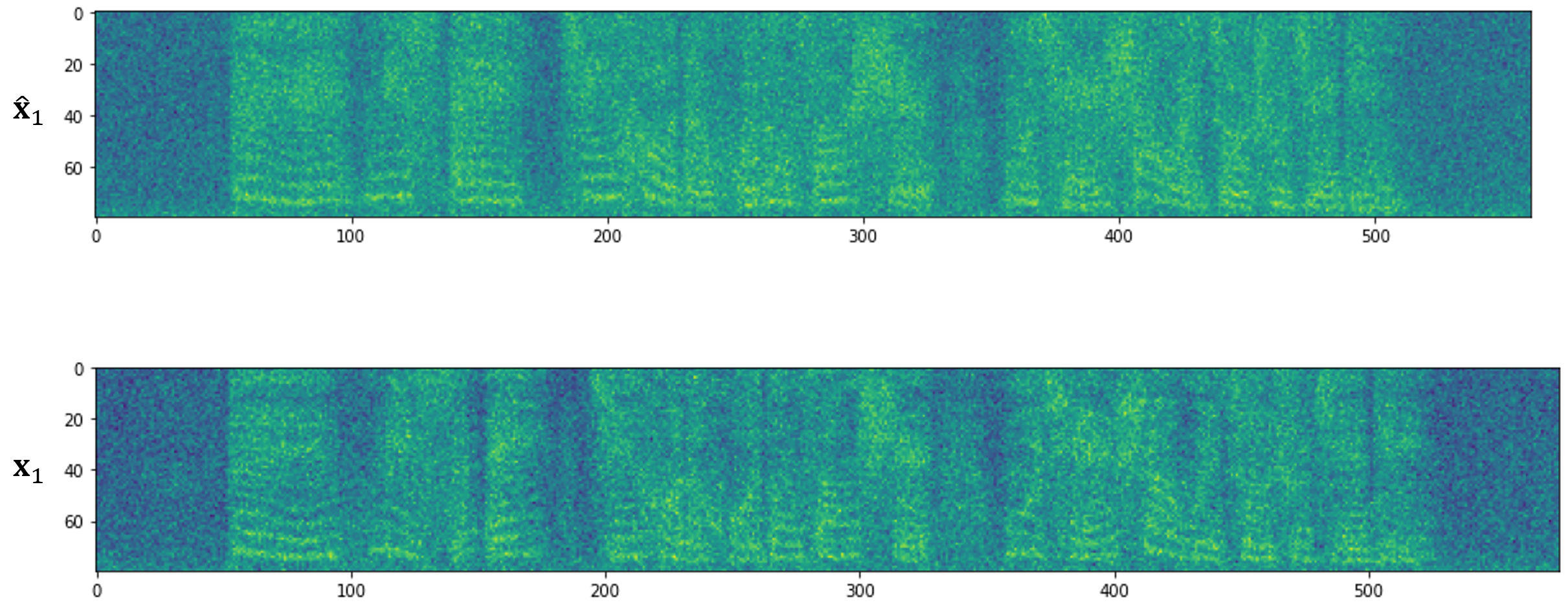}				
	\caption{Diffused samples at step 1. $\hat{\mathbf{x}}_0$: diffused sample from predicted mel spectrogram. $\mathbf{x}_0$: diffused sample from ground-truth mel spectrogram.}
	\label{fig:x1}
\end{figure}

\begin{figure}[t]
	\centering
	\includegraphics[width=0.45\textwidth]{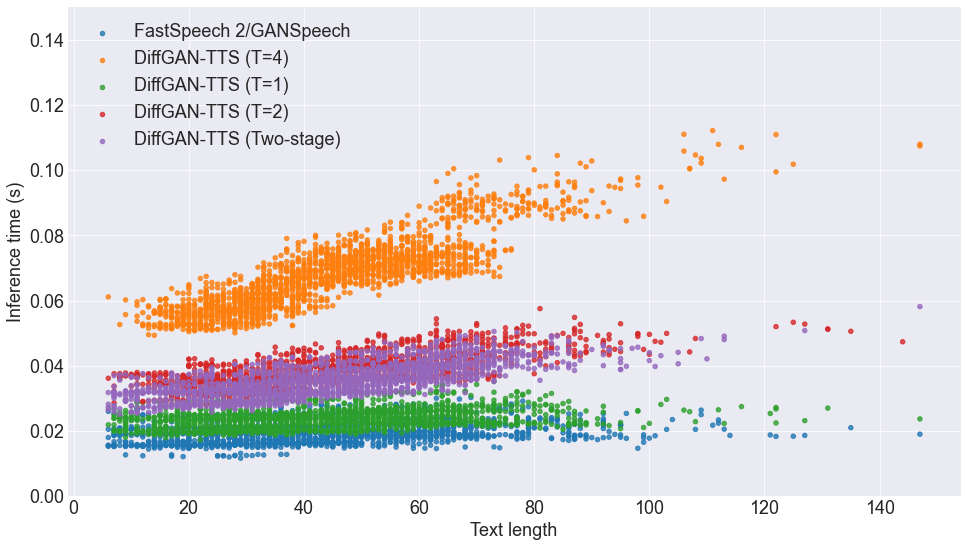}				
	\caption{Inference time (second) \textit{v.s.} text length (given in number of phonemes).}
	\label{fig:speed}
\end{figure}

We conduct crowd-sourced mean opinion score (MOS) tests to evaluate the quality of the generated speech perceptually. We keep the text content consistent among different models to exclude other interference factors, only examining the audio quality. Each audio sample is rated by at least 20 testers, who are asked to estimate the quality of synthesized speech on a nine-point Likert scale, with the lowest and highest scores being 1 (``Bad") and 5 (``Excellent"), and with an increment step of 0.5.
A subset of the audio samples is available online  \footnote{\url{https://anonym-demo.github.io/diffgan-tts/}}.
The evaluation results are shown in the last column of Table~\ref{tab:1}. We observe that the DiffSpeech model ($T$=60) outperforms other models in spite of its sub-optimal performance in objective measurement. This indicates that diffusion models may take a trade-off between model capacity and inference speed and be able to obtain higher quality with a larger number of denoising steps.
Overall, the DiffGAN-TTS family achieves good MOS results, with DiffGAN-TTS ($T$=4) obtaining the second-highest MOS (i.e., 4.22). We can see that DiffGAN-TTS (two-stage) achieves a MOS of 4.17, which is significantly better than DiffGAN-TTS ($T$=1 and 2). This verifies the two-stage training scheme and the active shallow diffusion mechanism introduced in Section~\ref{sec:active}. A visualization of taking one diffusion step on a predicted mel spectrogram by DiffGAN-TTS (two-stage) at training stage 1 (denoted as $\hat{\mathbf{x}}_1$) and its corresponding ground truth (denoted as $\mathbf{x}_1$) is shown in Figure~\ref{fig:x1}. We see that $\hat{\mathbf{x}}_1$ has a similar harmonic structure to $\mathbf{x}_1$, which inspires us to use the former as a strong prior for the latter.

\begin{table}[]
\caption{Objective metrics for ablation study.}
\vspace{0.3cm}
\centering
\resizebox{0.48\textwidth}{!}{
\begin{tabular}{ccccc}
\thickhline
Model               & SSIM      & MCD$_{24}$    & F$_0$ RMSE & Cos. Sim.              \\ \hline
DiffGAN-TTS         & \multirow{2}{*}{0.532} & \multirow{2}{*}{4.93} & \multirow{2}{*}{45.68} & \multirow{2}{*}{0.806} \\
($T$=4)               &                        &                       &                        &                        \\ \hline
$-\mathcal{L}_{fm}$ & 0.529  & 5.23 & 47.99 & 0.782                  \\
$-\mathcal{L}_{mel}$ & 0.492 & 5.86 & 48.97 & 0.710                   \\
$-\mathcal{L}_{fm}-\mathcal{L}_{mel}$ & \multicolumn{4}{c}{{[}Does not train{]}}                                                         \\
$+$ latent $\mathbf{z}$ & 0.517 & 5.10  & 46.68 & 0.794     \\ \thickhline
\end{tabular}}\label{tab:ablation}
\end{table}

\begin{figure}[t]
	\centering
	\includegraphics[width=0.42\textwidth]{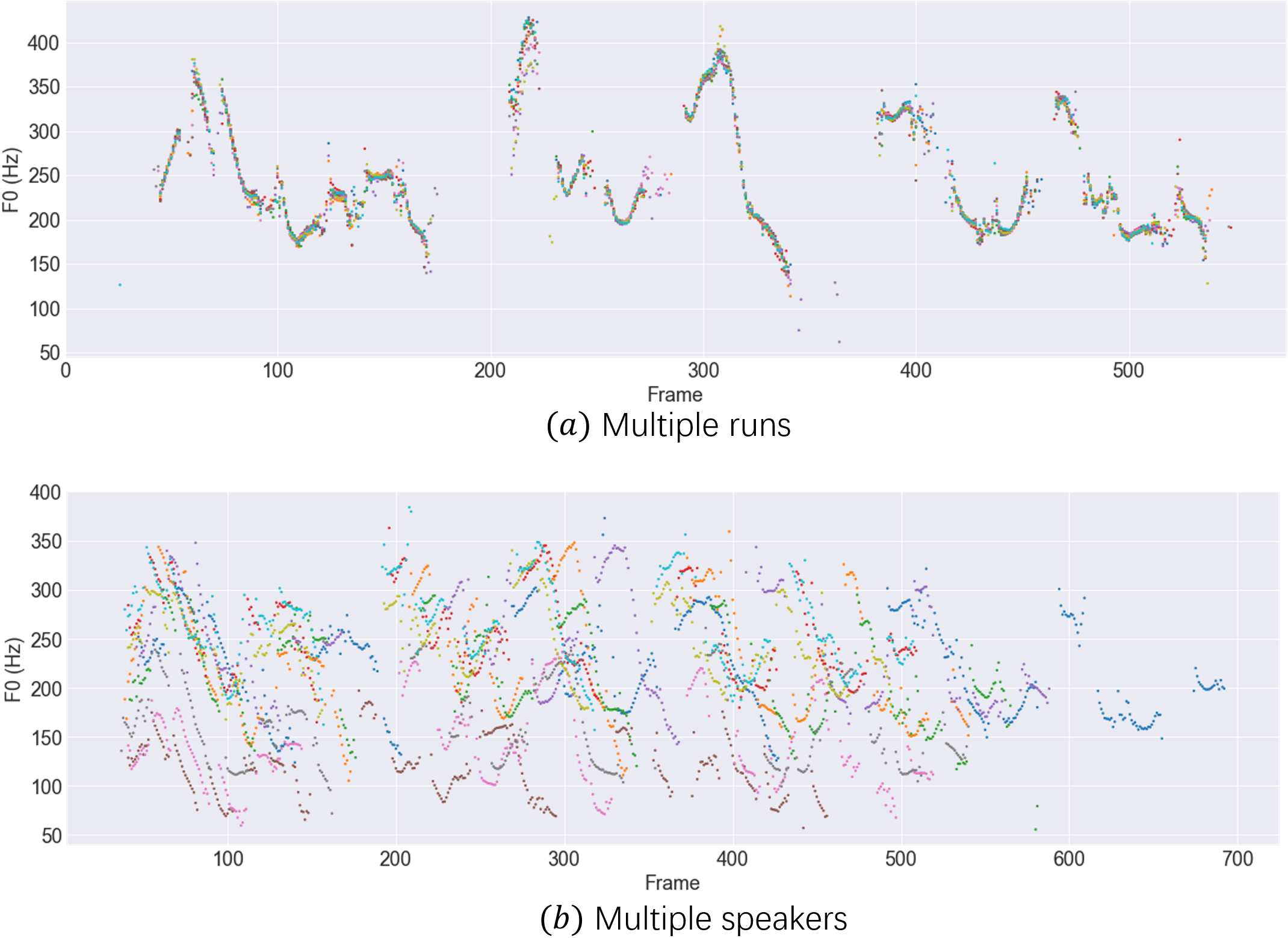}				
	\caption{Pitch tracks. ($a$): $F_0$ contours of 10 different runs with the same text input and speaker conditioning. ($b$): $F_0$ contours of 10 different speakers with the same text input.}
	\label{fig:diversity}
\end{figure}

\subsection{Synthesis Speed}

We assess the synthesis speed of mel spectrograms in terms of the Real-Time factor (RTF), indicating how many seconds it takes to generate one second of audio, on an NVIDIA T4 GPU and the number of model parameters. The efficiency information for all models is presented in Table~\ref{tab:1}. We also evaluate the scaling performance (inference time \textit{v.s.} text length), as illustrated in Figure~\ref{fig:speed}, where DiffSpeech is not depicted since its inference is too slow, making the plots of other models indistinguishable. It shows that DiffGAN-TTS ($T$=1) has similar scaling performance to FastSpeech 2 and other models have satisfactory scaling performance.

\subsection{Ablation studies}
\label{sec:ablation}

We conducted ablation studies in DiffGAN-TTS ($T$=4) to demonstrate the effectiveness of the use of mel loss $\mathcal{L}_{mel}$ and feature matching loss $\mathcal{L}_{fm}$ introduced in Section~\ref{sec:loss}, as well as the exclusion of modeling a latent variable $\mathbf{z}$ in the diffusion decoder. Details of how to use the latent variable are presented in Appendix~\ref{app:1}. Objective metrics are computed for these ablations. The results are shown in Table~\ref{tab:ablation}.
The model without using $\mathcal{L}_{mel}$ and $\mathcal{L}_{fm}$ does not train at all, which demonstrates that an adversarial loss alone is not sufficient for training a multi-speaker DiffGAN-TTS model. When comparing the model without $\mathcal{L}_{fm}$ to that without $\mathcal{L}_{mel}$, we observe that mel loss is more important than feature matching loss to successfully train a DiffGAN-TTS model. Adding a latent variable into the model makes all four metrics degrade more or less, indicating that the variance adaptor and speaker conditioning manage to model acoustic variations.

\subsection{Synthesis Variation}

Unlike the FastSpeech 2 and GANSpeech models, whose output is uniquely determined by the input text and speaker conditioning at inference, DiffGAN-TTS takes sampling processes at denoising steps and can inject some variations into the generated speech. To demonstrate this, we run a DiffGAN-TTS ($T$=4) model 10 times for a particular input text and speaker and compute the $F_0$ contours of the generated speech samples. We visualize in Figure~\ref{fig:diversity}($a$) and observe that DiffGAN-TTS generates speech with diverse pitches. We also generate speech samples for 10 speakers with the same text input, whose $F_0$ contours are visualized in Figure~\ref{fig:diversity}($b$), demonstrating that DiffGAN-TTS expresses very different prosody patterns for each speaker.

\section{Related Work}
\label{sec:related_work}

The Diffusion model is a family of generative models with the capacity to model complex data distribution and has attracted a lot of research attention in recent years.
Two streams of efforts have pushed research on diffusion models forward. One research stream is on score matching models \cite{JMLR:v6:hyvarinen05a}, where the problem of data density estimation is simplified into a score matching problem, i.e., estimating the gradient of the data distribution probabilistic density. The other stream is research on denoising diffusion probabilistic models (DDPMs) \cite{sohl2015deep}, where a diffusion Markov chain is adopted to add noise to the data structure and another Markov chain-based reverse process is used to generate data from noise. Lately, progress has been made in these two streams. The diffusion models have been shown to be able to generate high-quality images \cite{ddpm}. Afterward, DDPMs achieve high-quality audio generation \cite{diffwave,wavegrad} using the same parameterization introduced in \cite{ddpm}. On the other hand, score matching models have also been proven to generate high-resolution images by adopting a neural network to estimate the log gradient of data density. Slightly later, \cite{song2021scorebased} generalizes and improves previous work in score matching models through the lens of stochastic differential equations (SDEs) and shows that DDPMs and score matching models are special cases under this unified framework. In spite of wide and successful applications of diffusion models in different tasks, sampling from them is not efficient and often requires hundreds or even thousands of denoising steps, making them unsuited for real-time applications.

This work is inspired partially by \cite{diffgan}, which applies diffusion denoising GANs to accelerate inference in DDPMs for image synthesis, whereas we focus on high-fidelity and efficient text-to-speech (TTS) synthesis.
Furthermore, we develop a new model architecture and employ new losses to make the model suitable for TTS tasks that cannot be accomplished naturally.
The two-stage version of DiffGAN-TTS is closely related to DiffSpeech \cite{diffsinger}, which introduces a shallow diffusion scheme by training a boundary prediction to find the starting diffusion step at inference. Our work is much different:
1) We train the model to actively fuse the diffusion processes starting from the ground-truth mel spectrogram and the coarse mel spectrogram produced by a basic acoustic model.
2) Our model is adversarial trained and has a much faster inference speed than DiffSpeech.
3) We concentrate on the more difficult multi-speaker TTS tasks, whereas DiffSpeech concentrates on single-speaker TTS tasks.
4) Our model uses coarse model predictions as conditioning in the diffusion module, whereas DiffSpeech uses text encodings.

\section{Conclusion}
\label{sec:conclusion}

In this work, we have presented DiffGAN-TTS, a novel diffusion model-based non-AR TTS model able to achieve high-fidelity and efficient speech synthesis.
DiffGAN-TTS adopts an expressive model as a denoising function to approximate the true denoising distribution with adversarial training. Large denoising steps are allowed in DiffGAN-TTS, leading to faster inference.
We show with challenging multi-speaker TTS experiments that DiffGAN-TTS is able to generate high-fidelity speech samples with only \textit{\textbf{four}} denoising steps.
To further accelerate its inference process, we propose an active shallow diffusion mechanism and devise a two-stage training scheme to fully leverage prior knowledge from a basic acoustic model. Experiments show that DiffGAN-TTS is able to achieve high TTS performance with only \textit{\textbf{one}} denoising step.
We hope that DiffGAN-TTS will be used in many speech processing applications, especially those requiring real-time speech generation, to enjoy the powerful modeling capacity of diffusion models.
We also want to point out that even though the proposed model achieves decent TTS, it is still within the cascade of an acoustic model with a neural vocoder paradigm. Extending the model to support end-to-end text-to-waveform generation could be a possible direction for a shorter TTS pipeline and even better synthesis performance.

\nocite{langley00}

\bibliography{example_paper}
\bibliographystyle{icml2022}

\newpage
\appendix
\onecolumn

\section{Model details}
\label{app:1}

\begin{figure}[h]
	\centering
	\includegraphics[width=0.49\textwidth]{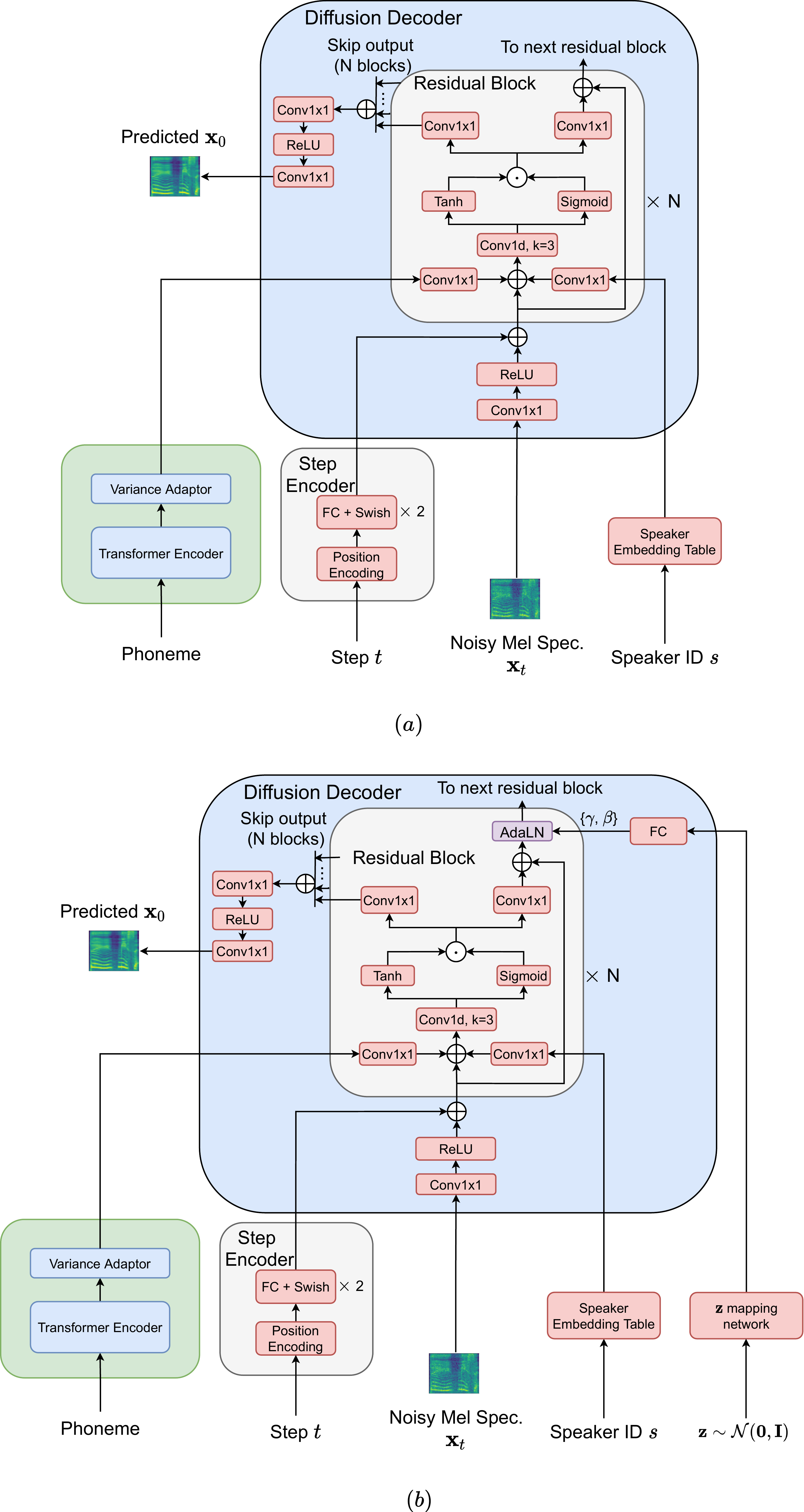}
	\caption{Model details: ($a$) Network structure of the diffusion decoder in DiffGAN-TTS models. ($b$) Network structure of the diffusion decoder modeling an extra latent variable $\mathbf{z}\sim\mathcal{N}(\mathbf{0},\mathbf{I})$.}
	\label{fig:model_detail}
\end{figure}

In this section, we present model details of the diffusion decoder in DiffGAN-TTS and the one in the ablation model taking an extra latent variable $\mathbf{z}$ as input (introduced in Section~\ref{sec:ablation}). The diffusion decoder has structure based on the residual block introduced in WaveNet \cite{wavenet}. Differently, we make the model non-causal. Figure~\ref{fig:model_detail}($a$) shows the diffusion decoder used in DiffGAN-TTS ($T$=1, 2 and 4) and as well as DiffGAN-TTS (two-stage). In the figure, \textit{FC} represents fully-connected layer, and \textit{Swish} represents the swish activation function \cite{swish}. Figure~\ref{fig:model_detail}($b$) shows the diffusion decoder where we inject a latent variable $\mathbf{z}\sim\mathcal{N}(\mathbf{0},\mathbf{I})$ as input. Inspired by StyleGAN \cite{stylegan} and following \cite{diffgan}, the latent variable $\mathbf{z}$ is first transformed by a mapping network, which is simply a fully-connected layer, into an embedding vector, denoted as $\mathbf{w}$. In each residual block, we incorporated the latent variable with adaptive layer normalization (AdaLN) layers. Specifically, a fully-connected layer is applied to convert $\mathbf{w}$ into scale $\mathbf{\gamma}$ and shift $\mathbf{\beta}$ features in each residual block. And then in each AdaLN layer, mean-variance normalization is first conducted on the residual output, and then we use $\mathbf{\gamma}$ and $\mathbf{\beta}$ to modulate the normalized feature $\mathbf{h}_{norm}$ as $\mathbf{h}_{norm} * \mathbf{\gamma} + \mathbf{\beta}$.  

\section{Training and inference algorithms}
\label{app:2}

\subsection{The Gaussian posterior}

In this section, we include a derivation for the Gaussian posterior distribution presented in \cite{ddpm} for completeness. Given a data point sampled from a real distribution $\mathbf{x}_0\sim q(\mathbf{x})$, consider the \textit{diffusion process} as a Markov chain as:
\begin{equation}
  q(\mathbf{x}_{1:T}|\mathbf{x}_0)=\prod_{t\geq 1}q(\mathbf{x}_t|\mathbf{x}_{t-1}), \quad q(\mathbf{x}_t|\mathbf{x}_{t-1}):=\mathcal{N}(\mathbf{x}_t; \sqrt{1-\beta_t}\mathbf{x}_{t-1},\beta_t\mathbf{I}).      
\end{equation}
Letting $\alpha_t=1-\beta_t$ and $\bar{\alpha}_t=\prod_{i=1}^t\alpha_i$, a nice property of the diffusion process is that we can sample $\mathbf{x}_t$ at any diffusion step $t$ in a closed form (a detailed derivation is given in Appendix A in \cite{diffwave}), as:
\begin{equation}\label{eq:forward_diffusion}
    q(\mathbf{x}_t|\mathbf{x}_0) = \mathcal{N}(\mathbf{x}_t;\sqrt{\bar{\alpha}_t}\mathbf{x}_0, (1-\bar{\alpha}_t)\mathbf{I}).
\end{equation}
By Bayes' rule and Markov chain property, the reverse conditional probability is tractable when conditioned on $\mathbf{x}_0$:
\begin{equation}\label{eq:cond_pos}
\begin{split}
    q(\mathbf{x}_{t-1}|\mathbf{x}_t, \mathbf{x}_0) &= q(\mathbf{x}_{t}|\mathbf{x}_{t-1}, \mathbf{x}_0)\frac{q(\mathbf{x}_{t-1}|\mathbf{x}_0)}{q(\mathbf{x}_t|\mathbf{x}_0)} \\
    & = \frac{q(\mathbf{x}_{t}|\mathbf{x}_{t-1})q(\mathbf{x}_{t-1}|\mathbf{x}_0)}{q(\mathbf{x}_t|\mathbf{x}_0)} \\
    & = \frac{\mathcal{N}(\mathbf{x}_t;\sqrt{\alpha_t}\mathbf{x}_{t-1},\beta_t\mathbf{I})\mathcal{N}(\mathbf{x}_{t-1};\sqrt{\bar{\alpha}_t}\mathbf{x}_0,(1-\bar{\alpha}_{t-1})\mathbf{I})}{\mathcal{N}(\mathbf{x}_{t};\sqrt{\bar{\alpha}_t}\mathbf{x}_0,(1-\bar{\alpha}_{t})\mathbf{I})} \\
    & \propto\exp{(-\frac{1}{2}(
    \frac{(\mathbf{x}_t-\sqrt{\alpha_t}\mathbf{x}_{t-1})^2}{\beta_t} + 
    \frac{(\mathbf{x}_{t-1}-\sqrt{\bar{\alpha}_{t-1}}\mathbf{x}_{0})^2}{1-\sqrt{\bar{\alpha}_{t-1}}} -
    \frac{(\mathbf{x}_t-\sqrt{\bar{\alpha}_t}\mathbf{x}_{0})^2}{1-\sqrt{\bar{\alpha}_t}}
    ))}  \\
    & =\exp{(
    -\frac{1}{2}((\frac{\alpha_t}{\beta_t}+\frac{1}{1-\bar{\alpha}_{t-1}})\mathbf{x}_{t-1}^2) - 
    (\frac{2\sqrt{\alpha_t}}{\beta_t}\mathbf{x}_t+\frac{2\sqrt{\bar{\alpha}_t}}{1-\sqrt{\bar{\alpha}_t}}\mathbf{x}_0)\mathbf{x}_{t-1} + C(\mathbf{x}_t,\mathbf{x}_0)
    )},
\end{split}
\end{equation}
where $C(\mathbf{x}_t,\mathbf{x}_0)$ is not involving $\mathbf{x}_{t-1}$. From the functional form in Eq.~\ref{eq:cond_pos} and some derivations, we observe that the posterior $q(\mathbf{x}_{t-1}|\mathbf{x}_t, \mathbf{x}_0)$ is a Gaussian distribution, which can be written as:
\begin{equation}\label{eq:q_sample}
    q(\mathbf{x}_{t-1}|\mathbf{x}_t, \mathbf{x}_0) = \mathcal{N}(
    \mathbf{x}_{t-1}; \tilde{\mathbf{\mu}}_t(\mathbf{x}_t, \mathbf{x}_0), \tilde{\beta}_t\mathbf{I}
    )
\end{equation}
with mean $\tilde{\mathbf{\mu}}_t(\mathbf{x}_t, \mathbf{x}_0)$ and variance $\tilde{\beta}_t$ having the forms of:
\begin{equation}
    \tilde{\mathbf{\mu}}_t(\mathbf{x}_t, \mathbf{x}_0)=
    \frac{\sqrt{\bar{\alpha}_{t-1}}\beta_t}{1-\bar{\alpha}_t}\mathbf{x}_0 + 
    \frac{\sqrt{\alpha_t}(1-\bar{\alpha}_{t-1})}{1-\bar{\alpha}_t}\mathbf{x}_t 
    \quad \text{and} \quad
    \tilde{\beta}_t=\frac{1-\bar{\alpha}_{t-1}}{1-\bar{\alpha}_t}\beta_t.
\end{equation}

\subsection{Variance schedule}

We use the discretization of the continuous-time extension of the diffusion process in Eq.~\ref{eq:forward_diffusion} with the variance preserving (VP) SDE \cite{song2021scorebased} to compute the variance schedule $\beta_1, \cdots, \beta_T$. Specifically, for $t\in\{1,\cdots,T\}$, we compute $\beta_t$ as:
\begin{equation}
    \beta_t = 1 - \exp(-\frac{\beta_{\text{min}}}{T} - 0.5(\beta_{\text{max}} - \beta_{\text{min}})\frac{2t-1}{T^2}),
\end{equation}
where we set the constants as $\beta_{\text{min}}=0.1$ and $\beta_{\text{min}}=40$ in all experiments.

\subsection{Training algorithm of DiffGAN-TTS}

The training procedures of the DiffGAN-TTS models with $T$=1, 2 and 4 are the same across our experiments. We summarize the algorithm in Algorithm~\ref{alg:1}.

\subsection{Inference algorithm of DiffGAN-TTS}

\begin{figure}[h]
	\centering
	\includegraphics[width=0.6\textwidth]{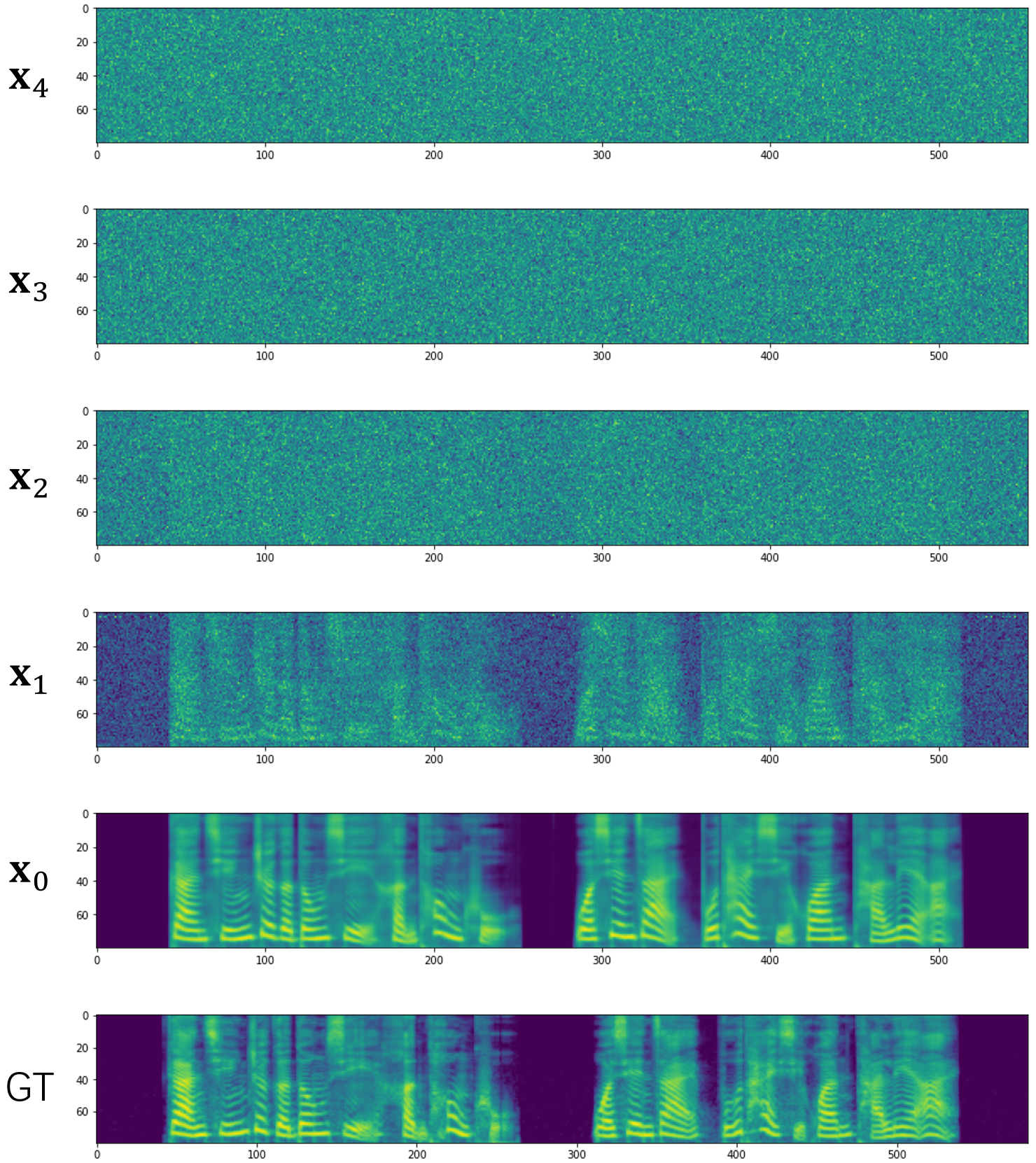}
	\caption{A visualization of the denoising steps in DiffGAN-TTS ($T$=4) at inference. ``GT" represents the ground-truth mel spectrogram.}
	\label{fig:mel_gen1}
\end{figure}

The inference procedures of the DiffGAN-TTS models with $T$=1, 2 and 4 are the same across our experiments. We summarize the algorithm in Algorithm~\ref{alg:2}.
We visualize the denoising steps of the DiffGAN-TTS ($T$=4) model in Figure~\ref{fig:mel_gen1}.

\subsection{Training algorithm of DiffGAN-TTS with active shallow diffusion}

\begin{figure}[h]
	\centering
	\includegraphics[width=0.6\textwidth]{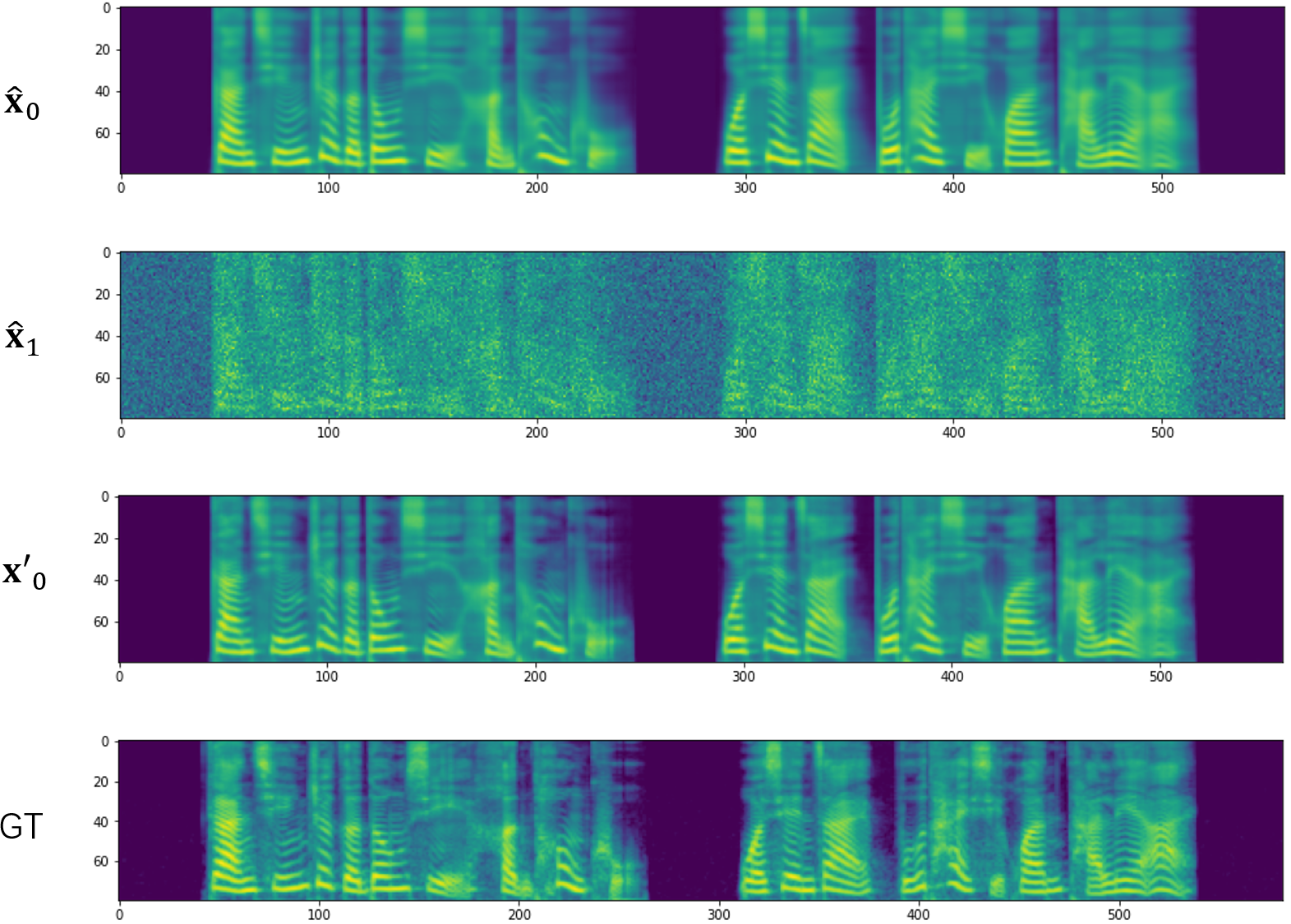}
	\caption{A visualization of the denoising steps in DiffGAN-TTS (two-stage) as inference, which uses the active shallow diffusion mechanism. ``GT" represents the ground-truth mel spectrogram.}
	\label{fig:mel_gen2}
\end{figure}

The training procedures of the DiffGAN-TTS model with active shallow diffusion, i.e., DiffGAN-TTS (two-stage), is summarized in Algorithm~\ref{alg:3}.
We visualize the denoising steps of the DiffGAN-TTS (two-stage) model in Figure~\ref{fig:mel_gen2}.

\subsection{Inference algorithm of DiffGAN-TTS with active shallow diffusion}

The inference procedures of the DiffGAN-TTS model with active shallow diffusion, i.e., DiffGAN-TTS (two-stage), is summarized in Algorithm~\ref{alg:4}.

\begin{algorithm}[h]
  \caption{Training procedure of DiffGAN-TTS}
  \label{alg:1}
\textbf{Input}: The acoustic generator $G_\theta$ with parameters $\theta$. The discriminator $D_\phi$ with parameter $\phi$. A pre-calculated variance schedule $\beta_1, \cdots,\beta_T$ with $T$ diffusion steps. The training set $\mathcal{D}_{train}=\{(\mathbf{x}_0, \mathbf{y}, \mathbf{s}, \mathbf{d},\mathbf{p}, \mathbf{e})_i\}_{i=1}^{N}$ with $N$ data points.
\begin{algorithmic}[1]
  \STATE Initialize parameters $\theta$ and $\phi$;
  \WHILE{\textit{model not converge}}
  \STATE Sample $M$ data points from $\mathcal{D}_{train}$ to form a mini-batch  $\mathcal{S}=\{(\mathbf{x}_0, \mathbf{y}, \mathbf{s}, \mathbf{d},\mathbf{p}, \mathbf{e})_i\}_{j=1}^{M}$;
  \FOR{\textit{data point} in $\mathcal{S}$}
  \STATE Uniformly sample $t$ from $[0,\cdots, T]$;
  \STATE Sample $\mathbf{x}_t$ and $\mathbf{x}_{t-1}$ given $\mathbf{x}_0$ according to Eq.\ref{eq:forward_diffusion};
  \STATE Add $t$, $\mathbf{x}_t$ and $\mathbf{x}_{t-1}$ to current data point;
  \ENDFOR
  \STATE Step I: train the discriminator
  \STATE Forward-propagate $\mathcal{S}$ to $G_\theta$ to obtain output $\mathcal{O}=\{(\mathbf{x}'_0, \hat{\mathbf{d}},\hat{\mathbf{p}}, \hat{\mathbf{e}})_i\}_{j=1}^{M}$;
  \STATE Sample the posterior $\mathbf{x}'_{t-1}$ for each $\mathbf{x}'_0$ in $\mathcal{O}$ by Eq.~\ref{eq:q_sample};
  \STATE Compute $\mathcal{L}_D$ by Eq.~\ref{eq:d-loss};
  \STATE Do back-propagation with $\mathcal{L}_D$ and update $\phi$ one step with gradient descent;
  \STATE Step II: train the acoustic generator
  \STATE Compute $\mathcal{L}_G$ by Eq.~\ref{eq:g-loss};
  \STATE Do back-propagation with $\mathcal{L}_G$ and update $\theta$ one step with gradient descent;
  \ENDWHILE
  \STATE {\bfseries Return} $G_{\theta}$;
\end{algorithmic}
\end{algorithm}

\begin{algorithm}[h]
  \caption{Inference procedure of DiffGAN-TTS}
  \label{alg:2}
  \textbf{Input}: A trained acoustic generator $G_\theta$ and one testing sample $(\mathbf{y}, s)$.
\begin{algorithmic}[1]
  \STATE Sample $\mathbf{x}_T\sim\mathcal{N}(\mathbf{0}, \mathbf{I})$;
  \STATE $\mathbf{x}_t \gets \mathbf{x}_T$;
  \FOR{$t=T,T-1,...,1$}
    \STATE Forward-propagate $(\mathbf{x}_t, \mathbf{y}, s, t)$ to $G_\theta$ to calculate $\mathbf{x}'_0$;
    \STATE Sample $\mathbf{x}'_{t-1}$ given $\mathbf{x}'_0$ and $\mathbf{x}_t$ by Eq.~\ref{eq:q_sample};
    \STATE $\mathbf{x}_t \gets \mathbf{x}'_{t-1}$;
  \ENDFOR
  \STATE {\bfseries Return} $\mathbf{x}_t$;
\end{algorithmic}
\end{algorithm}

\begin{algorithm}[h]
  \caption{Training procedure of DiffGAN-TTS with active shallow diffusion}
  \label{alg:3}
\textbf{Input}: The basic acoustic model $G_{\psi}^{\text{base}}$ with parameters $\psi$. The acoustic generator $G_\theta$ with parameters $\theta$. The discriminator $D_\phi$ with parameter $\phi$. A pre-calculated variance schedule $\beta_1, \cdots,\beta_T$ with $T$ diffusion steps. The training set $\mathcal{D}_{train}=\{(\mathbf{x}_0, \mathbf{y}, \mathbf{s}, \mathbf{d},\mathbf{p}, \mathbf{e})_i\}_{i=1}^{N}$ with $N$ data points. Number of basic acoustic model training iterations $N_{\text{iter}}$.
\begin{algorithmic}[1]
  \STATE Initialize parameters $\psi$, $\theta$ and $\phi$;
  \STATE {\bfseries Stage 1:} train the basic acoustic model;
  \FOR{$i=1,2,...,N_{\text{iter}}$}
  \STATE Sample $M$ data points from $\mathcal{D}_{train}$ to form a mini-batch  $\mathcal{S}=\{(\mathbf{x}_0, \mathbf{y}, \mathbf{s}, \mathbf{d},\mathbf{p}, \mathbf{e})_i\}_{j=1}^{M}$;
  \STATE Forward-propagate $\mathcal{S}$ to $G_{\psi}^{\text{base}}$ to obtain output $\mathcal{O}=\{(\hat{\mathbf{x}}_0, \hat{\mathbf{d}},\hat{\mathbf{p}}, \hat{\mathbf{e}})_i\}_{j=1}^{M}$;
  \STATE Compute reconstruction loss $\mathcal{L}_{recon}$ by Eq.~\ref{eq:reconst-loss};
  \STATE Do back-propagation with $\mathcal{L}_{recon}$ and update $\psi$ one step with gradient descent;
  \ENDFOR
  \STATE {\bfseries Stage 2:} train the diffusion encoder in $G_\theta$;
  \STATE Initialize partial weights in $\theta$ with $\psi$ as introduced in Section~\ref{sec:active};
  \WHILE{\textit{model not converge}}
  \STATE Sample $M$ data points from $\mathcal{D}_{train}$ to form a mini-batch  $\mathcal{S}=\{(\mathbf{x}_0, \mathbf{y}, \mathbf{s}, \mathbf{d},\mathbf{p}, \mathbf{e})_i\}_{j=1}^{M}$;
  \FOR{\textit{data point} in $\mathcal{S}$}
  \STATE Uniformly sample $t$ from $[0,\cdots, T]$;
  \STATE Sample $\mathbf{x}_t$ and $\mathbf{x}_{t-1}$ given $\mathbf{x}_0$ according to Eq.\ref{eq:forward_diffusion};
  \STATE Add $t$, $\mathbf{x}_t$ and $\mathbf{x}_{t-1}$ to current data point;
  \ENDFOR
  \STATE Step I: train the discriminator
  \STATE Forward-propagate $\mathcal{S}$ to $G_\theta$ to obtain output $\mathcal{O}=\{(\mathbf{x}'_0, \hat{\mathbf{d}},\hat{\mathbf{p}}, \hat{\mathbf{e}})_i\}_{j=1}^{M}$;
  \STATE Sample the posterior $\mathbf{x}'_{t-1}$ for each $\mathbf{x}'_0$ in $\mathcal{O}$ by Eq.~\ref{eq:q_sample};
  \STATE Compute $\mathcal{L}_D$ by Eq.~\ref{eq:d-loss};
  \STATE Do back-propagation with $\mathcal{L}_D$ and update $\phi$ one step with gradient descent;
  \STATE Step II: train the acoustic generator
  \STATE Compute $\mathcal{L}_G$ by Eq.~\ref{eq:g-loss};
  \STATE Do back-propagation with $\mathcal{L}_G$ and update $\theta$ one step with gradient descent;
  \ENDWHILE
  \STATE {\bfseries Return} $G_{\theta}$;
\end{algorithmic}
\end{algorithm}

\begin{algorithm}[h]
  \caption{Inference procedure of DiffGAN-TTS with active shallow diffusion using \textbf{\textit{one}} denoising step}
  \label{alg:4}
  \textbf{Input}: A trained acoustic generator $G_\theta$ and one testing sample $(\mathbf{y}, s)$.
\begin{algorithmic}[1]
  \STATE Forward $(\mathbf{y}, s)$ through transformer encoder, variance adaptor and mel decoder to obtain coarse prediction $\hat{\mathbf{x}}_0$;
  \STATE Sample $\hat{\mathbf{x}}_1$ conditioning on $\hat{\mathbf{x}}_0$ by Eq.~\ref{eq:forward_diffusion};
  \STATE Forward-propagate $(\hat{\mathbf{x}}_1, y, s, t=1)$ to the diffusion decoder in $G_\theta$ to calculate $\mathbf{x}'_0$;
  \STATE {\bfseries Return} $\mathbf{x}'_0$;
\end{algorithmic}
\end{algorithm}

\end{document}